\documentclass[preprint]{revtex4-2}
\usepackage{siunitx}
\usepackage{graphicx}
\usepackage{latexsym}

\draft 

\begin{document}

\title{Structural Analysis of Amorphous GeO$_2$ under High Pressure Using Reverse Monte Carlo Simulations}



\author{Kenta Matsutani}
\affiliation{Graduate School of Science and Engineering, Yamagata university, 1-4-12, Kojirakawa, Yamagata 990-8560, Japan}
\author{Asumi Yamauchi}
\affiliation{Faculty of Science, Yamagata University, 1-4-12 Kojirakawa, Yamagata 990-8560, Japan}
\author{Shusuke Kasamatsu}
\affiliation{Faculty of Science, Yamagata University, 1-4-12 Kojirakawa, Yamagata 990-8560, Japan}
\author{Takeshi Usuki}
\affiliation{Faculty of Science, Yamagata University, 1-4-12 Kojirakawa, Yamagata 990-8560, Japan}

\date{\today}

\begin{abstract}
The structural properties of amorphous GeO$_2$, a prototypical network glass, were investigated under ambient to high pressure using reverse Monte Carlo simulations based on reported structure factors from in situ high-pressure neutron diffraction experiments with isotopic substitution. 
The results indicate the retention of the topological structure containing predominantly tetrahedral GeO$_4$ units up to ca.~\SI{4}{\giga\pascal} ($\rho/\rho_0 = 1.15$), which is explained by the reduction of cavity volumes. With further application of pressure, an increase in the number of GeO$_5$ units is first observed, which is then followed more gradually by an increase in the number of GeO$_6$ units.

\end{abstract}

\maketitle 

\section{I\lowercase{ntroduction}}
The structural compaction mechanism of vitreous oxides under high pressure has been a topic of high interest in the context of geoscience with the ultimate aim of understanding the nature of Earth's interior \cite{lee2008, petitgirard2019}. The topic is also of significance from the standpoint of glass theory; characterizing and understanding the nature of pressure-driven transitions in glassy systems is still a difficult and fundamental challenge \cite{wilding2006, onodera2020}. 
A prototypical oxide system under intense study in this regard is GeO$_2$. Although its chemical (and to a large extent, structural) analogue SiO$_2$ is known for more practical applications, GeO$_2$ is an ideal system for studying structural transition of vitreous oxides because of its higher sensitivity to pressure \cite{micoulaut_structure_2006}. 

At ambient pressure, amorhpous GeO$_2$ is comprised mostly of a network of corner-sharing GeO$_4$ tetrahedral units. Experimental \cite{itie1989,lelong2012,wezka2012} and theoretical \cite{shanavas2006,li2009,wezka2012,zhu2009,brazhkin2011} reports in the literature basically agree that this changes gradually to a network consisting of predominantly octahedral GeO$_6$ units with increasing pressure. A similar behavior is known for crystalline GeO$_2$ although the transition occurs much more abruptly \cite{itie1989,vaccari2009,brazhkin2011}. 
There are some discrepancies among the literature on the details of the transition process, however.
For example, Lelong et al.~conducted inelastic X-ray scattering (IXS) experiments on compressed amorphous GeO$_2$ and used crystalline standards to estimate the coordination environment of Ge atoms \cite{lelong2012}. They report that five-fold coordinated Ge occurs as a minor component (less than \SI{30}{\percent}) only in the range of \SIrange{5}{8}{\giga\pascal} and acts as a transient species between four- and six-fold coordinated Ge. On the other hand, Wezka et al.~have presented a different picture based on a combination of molecular dynamics (MD) simulations employing their original DIPole-Polarizable Ion Model (DIPPIM) \cite{marrocchelli2009,marrocchelli2010} and {\it in situ} neutron diffraction with isotopic substituion (NDIS) \cite{salmon2007,wezka2012}. They first show that DIPPIM reproduces almost perfectly the pressure dependence of the average bond lengths and coordination numbers obtained from NDIS when compared at the same densities (note that MD fails to reproduce the pressure-density relationship, most likely due to the limited timescale in the glass preparation procedure), then go on to interpret local non-average information from MD. The DIPPIM MD results show that the tetrahedral network is sustained up to $\sim$\SI{4.5}{\giga\pascal}, after which five-fold coordinated Ge starts to increase. Six-fold coordinated Ge also start to appear above $\sim$\SI{8}{\giga\pascal}, but no suppression of five-fold coordinated Ge appears in contrast to IXS results. It was postulated that the discrepancies stem from using crystalline standards containing trigonal bipyramidal GeO$_5$ which was found to be rather scarce in DIPPIM results \cite{wezka2012}. Various other classical and {\it ab initio} MD studies can be found in the literature, and the overall behavior is reported to be rather similar to DIPPIM although with some quantitative differences \cite{shanavas2006,li2009,zhu2009,brazhkin2011}.

One information that is now lacking is the experimental confirmation of the evolution of the ratio of coordination environments vs.~pressure, since neutron diffraction only provides direct information on the \emph{average} coordination and not the local ratios. Also, even though the MD simulation using DIPPIM shows excellent agreement in various average quantities available from NDIS, the agreement of the structure factors is far from perfect, especially at higher pressures \cite{wezka2012}.
The present study addresses this issue through reverse Monte Carlo (RMC) modeling \cite{evrard2005},
which attempts to obtain a real-space structure that reproduces the structure factors obtained by the diffraction experiment. 
A well known issue
with RMC modeling based on total structure factors is that they sometimes result in rather ``unphysical''
results with many 
homopolar bonds, because partial correlations cannot be deduced from total structure factors.
This issue is alleviated to a large extent by using structure factors measured with different isotopic enrichments, i.e., NDIS,
which allows for reproduction of the correct partial (Ge--O, Ge--Ge, and O--O) correlations \cite{salmon2007}.
Here, we perform RMC modeling 
using NDIS data below \SI{8}{\giga\pascal} and without isotopic substitution up to \SI{17.5}{\giga\pascal}.
Coordination numbers and bond angles extracted from the obtained RMC model are compared carefully with previous experimental and computational literature. To complement the analysis, we also performed FPMD simulations of glass formation via a melt-quench approach.

\section{M\lowercase{odeling} P\lowercase{rocedure}}
RMC modeling of amorphous GeO$_2$ was conducted using RMC++ code \cite{evrard2005}.
We employed models containing 3000 atoms (Ge 1000 and O 2000 atoms, respectively) in cubic simulation cells under periodic 
boundary conditions.
The starting atomic configurations corresponding to experimental number density at ambient pressure  (\SI{0.0629}{\angstrom^{-3}} \cite{leadbetter1972}) were generated by a hard-sphere Monte Carlo simulation. 
Then, RMC modeling was performed based on structure factors \textit{F}(\textit{Q}) from high-pressure NDIS \cite{salmon2007} from ambient to \SI{8}{\giga\pascal}, and without isotopic substitution 
at \SIrange{8.5}{17.5}{\giga\pascal} ($1.38 < \rho/\rho_0 < 1.64$) \cite{salmon2012}. 
The number density was raised in steps of 0.0629, 0.0727, 0.082, 0.0845, 0.0867, 0.0868, 0.0951, 0.0987, and 0.1031 \r{A}$^{-3}$ to simulate pressure application.
The corresponding experimental pressure values calculated by interpolation of experimental pressure-volume data are 4, 5.9, 6.8, 8.0, 8.5, 11.5, 14.5, \SI{17.5}{\giga\pascal}.
The final structure obtained by RMC simulation at each pressure was taken to be the initial structure for the next pressure step.
Coordination constraints are used to avoid two- or three-fold coordinated Ge; we judged such coordination environments to be
very scarce and `unphysical' based on FPMD results, where no Ge atoms were found with less than four-fold coordination.
The cut-off radii used in RMC modeling are tabulated below. (table 1)
We detected cavities in the models by determining domains that are farther from any atoms than a cutoff distance of \SI{2.8}{\angstrom} using Voronoi construction as implemented in pyMolDyn \cite{heimbach2017}.

\begin{table}[hbtp]
    \caption{The cutoff radius at each pressure controls the closest distance between two atoms in a glass system.}
    \label{tab:my_label}
    \centering
    \begin{tabular}{cccc}
        \hline
        Pressure  (GPa) & Ge-Ge (\si{\angstrom}) & Ge-O (\si{\angstrom}) & O-O (\si{\angstrom})\\
        \hline \hline
        ambient & 2.6 & 1.6 & 2.45\\
        4.0 & 2.6 & 1.6 & 2.4\\
        5.9 & 2.6 & 1.6 & 2.3\\
        6.8 & 2.6 & 1.6 & 2.3\\
        8.0 & 2.6 & 1.6 & 2.3\\
        8.5 & 2.7 & 1.65 & 2.2\\
        11.5 & 2.8 & 1.7 & 2.2\\
        14.5 & 2.625 & 1.6 & 2.2\\
        17.5 & 2.65 & 1.6 & 2.2\\
        \hline
    \end{tabular}
\end{table}

FPMD simulations for amorphous GeO$_2$ were performed with the generalized gradient approximation functional by Perdew-Burke-Ernzerhof (GGA-PBE) \cite{perdew1996} implemented in Vienna ab initio Simulation Package (VASP) code \cite{kresse1996}. 
The projector augmented wave method was used to describe electron-ion interactions. 
The plane wave cutoff energy was set at \SI{289.8}{\electronvolt}, and the self-consistent-field convergence criteria were adopted to be smaller than $10^{-5}$ eV. 
Only the $\Gamma$ point is used in the Brillouin zone sampling.
To obtain the glass structure, we first randomly placed 120 atoms (40 Ge atoms and 80 O atoms) in a cubic supercell while keeping
all atoms at least \SI{1.8}{\angstrom} apart. The cell size was determined so that the number density corresponds to
the experimental one for the amorphous phase at ambient pressure.
Using the Nos\'e thermostat, the model was heated up to \SI{3000}{\kelvin} and equilibrated for \SI{40}{\pico\second} in the
liquid state. 
Subsequently, the model was quenched to \SI{300}{\kelvin} with a rate of \SI{5.0}{\kelvin/\pico\second}, then equilibrated for \SI{40}{\pico\second} to yield the FPMD-derived structure for amorphous GeO$_2$ under ambient conditions.

\section{R\lowercase{esults and} D\lowercase{iscussion}}
\subsection{Structure Factors}
\begin{figure}[ht]
    \centering
    \includegraphics[width=\columnwidth]{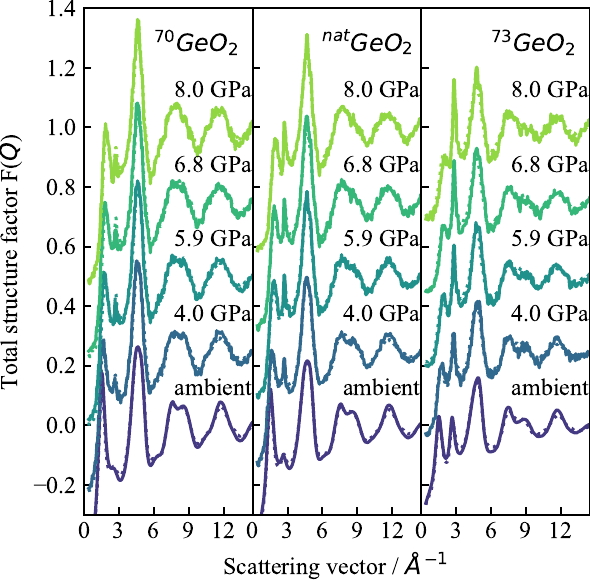}
    \caption{(Color online) The pressure dependence of total structure factor of amorphous $^{70}$GeO$_2$, $^{nat}$GeO$_2$ and $^{73}$GeO$_2$. The solid curves give the experimental data, and the dotted curves represent RMC results.}
    \label{fig:sf}
\end{figure}

First, we compare RMC and experimental structure factors in order to confirm our RMC analysis
reproduces the structure factor from neutron diffraction with isotope substitution.
Figure \ref{fig:sf} shows that our RMC model successfully fits all three experimental total structure
factors of samples isotopically enriched with $^{70}$Ge, $^{73}$Ge, 
and that using natural isotopic abundance of Ge denoted by $^\text{nat}$Ge.
Each total structure factor function has the first sharp diffraction peak (FSDP) at \SI{1.5}{\angstrom^{-1}}, 
which decreases its magnitude upon compression. 
On the other hand, the small second peak (principal peak) at ambient pressure increases in intensity with application of pressure; this indicates that pressure application leads to changes in the intermediate-range structure.
Thus, we can be fairly certain that our RMC model contains the correct partial correlations.

\begin{figure*}
    \centering
    \includegraphics[width=15.5cm]{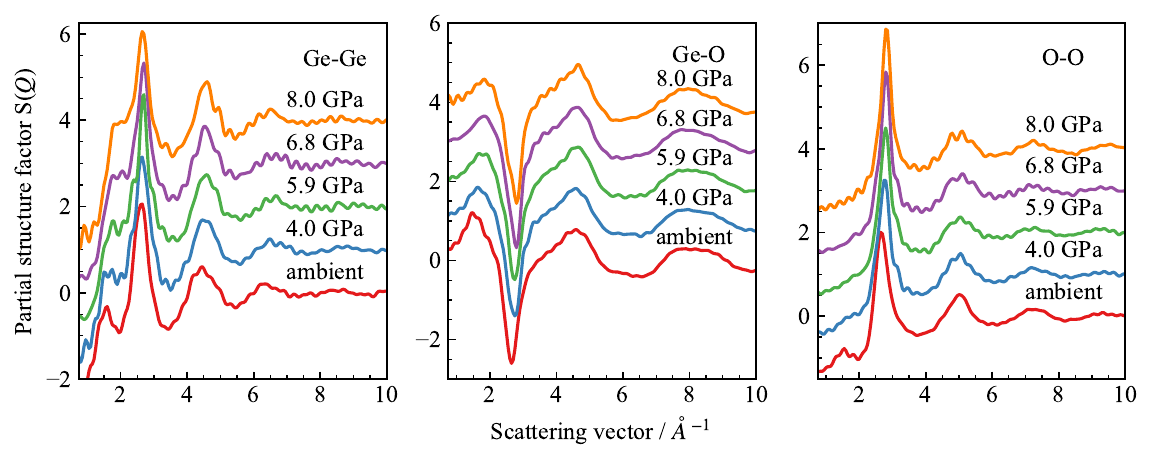}
    \caption{(Color online) The pressure dependence of partial structure factors for GeO$_2$ glass. The partial structure factors at ambient pressure, \SI{4.0}{\giga\pascal}, \SI{5.9}{\giga\pascal}, \SI{6.8}{\giga\pascal}, and \SI{8.0}{\giga\pascal} are stacked from bottom to top.}
    \label{fig:p_sq}
\end{figure*}

To further understand the details of structural changes in GeO$_2$ glass, we examine the partial structure factors shown in figure \ref{fig:p_sq}.
The pressure-induced changes in partial structure factors show similar features to the total structure factors.
At ambient pressure, a FSDP is observed in all partial structure factors. The Ge--Ge and Ge--O FSDPs show similar behavior upon pressure application: the intensities are suppressed and the positions are shifted to higher \textit{Q} values, finally merging with the principal peak at above \SI{8}{\giga\pascal}.
On the other hand, the O--O FSDP is completely suppressed at \SI{4}{\giga\pascal} and above. This correlates well with the annihilation of cavities as will be discussed in Sec. III.D. The principal peak in the O--O partial structure factor also shifts to higher \textit{Q} values, but the peak intensity is increased in contrast to the FSDP. This behavior contrasts with that observed for SiO$_2$ glass [23], as the O-O partial structure factors in SiO$_2$ glass are almost unchanged under \SI{6}{\giga\pascal}. This is a manifestation of the higher sensitivity of GeO$_2$ glass structure to pressure compared to SiO$_2$.

\subsection{Coordination Number}
\begin{figure*}
    \centering
    \includegraphics[width=15cm]{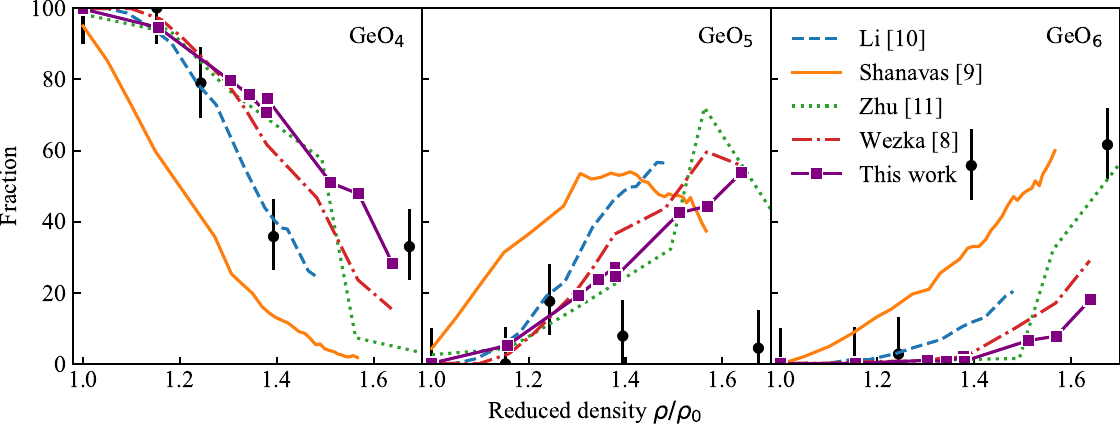}
    \caption{(Color online) The reduced density dependence of the fraction of GeO$_4$ (left), GeO$_5$ (middle) and GeO$_6$ (right) units in amorphous GeO$_2$.
    The present RMC results (solid curves with squares) are compared to molecular dynamics results in the literature \cite{li2009,shanavas2006,zhu2009,wezka2012}. 
    The fractions of GeO$_x$ ($x=4, 5, 6$) units deduced from IXS experiments \cite{lelong2012} are also shown ($\bullet$ with error bars $\pm$\SI{10}{\percent}).}
    \label{fig:coord}
\end{figure*}
Coordination number changes around germanium atoms are representative of short-range structural order. 
The importance of these features have recently been recognized especially in terms of geoscience \cite{guthrie2004, sanloup2013}.
Figure \ref{fig:coord} shows the fraction of GeO$_x$ ($x=4, 5, 6$) units calculated from the present RMC results as a function of reduced density and compares them to classical and {\it ab initio} MD studies and IXS experimental results \cite{li2009,shanavas2006,wezka2012,zhu2009,lelong2012}.
Our RMC model shows retention of the tetrahedral structure up to $\rho/\rho_0 \sim 1.15$ after which the number of 
GeO$_4$ units start to decrease and the number of GeO$_5$ units start to increase. This signifies the conversion of 
tetrahedral GeO$_4$ units to square pyramidal GeO$_5$ units and
is generally in good agreement with the literature. One exception is the MD work by Shanavas {\it et al.} \cite{shanavas2006}
where no retention is seen and structural transitions occur immediately upon application of pressure.
With further compaction, the number of GeO$_5$ units increase continuously, reaching \SI{50}{\percent} at 
$\rho/\rho_0=1.64$.
Octahedral GeO$_6$ units start to emerge more gradually than GeO$_5$ at about $\rho/\rho_0 = 1.38$, 
indicating that GeO$_5$ units act as precursors for conversion to GeO$_6$. 
Again, the general trend is in good agreement with most of the literature.
A clear exception is the IXS result \cite{lelong2012}, which shows emergence of a small amount of 
GeO$_5$ units at $\rho/\rho_0 \sim 1.25$ followed by its suppression and a stepwise increase in the number 
of GeO$_6$ units.
The IXS result is very similar to pressure-induced structural transitions observed in the crystalline phase;
thus, Wezka and coworkers \cite{wezka2012} suggested that the disagreement stems from using crystalline standards \cite{lelong2012}, 
which included trigonal bipyramids as GeO$_5$ units \cite{cabaret2007}, to analyze IXS results.

\subsection{Bond Angle Distributions}
\begin{figure}
    \centering
    \includegraphics{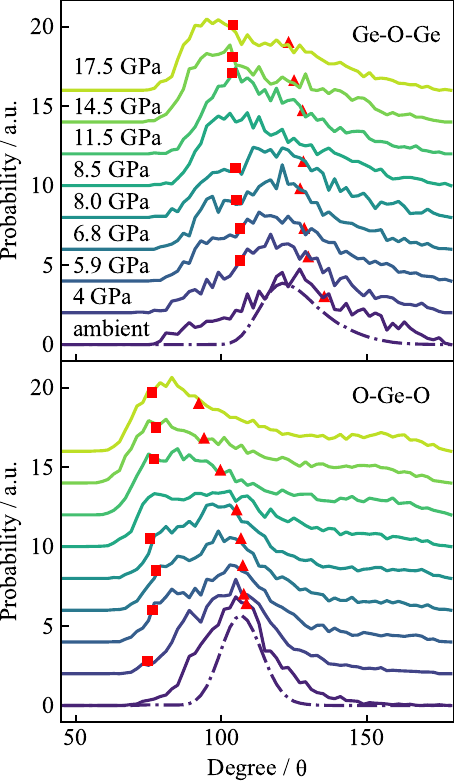}
    \caption{\label{fig:angle}(Color online) Ge--O--Ge (upper) and O--Ge--O (bottom) bond angle distributions calculated from our RMC models
    (solid lines) at each pressure, as well as our ambient pressure FPMD result (chained line). 
    The results are compared to bond angles
    (peaks and shoulders in the bond angle distributions) from classical MD with DIPPIM potentials, where $\triangle$
    correspond to GeO$_4$ motifs and $\Box$ correspond to GeO$_5$ and GeO$_6$ units.
    Note the intensity of our FPMD results was divided by 2 for clarity.
    }
\end{figure}

Next, the Ge--O--Ge and O--Ge--O bond angle distributions, which are often used to understand intermediate-range order beyond nearest neighbor atoms, were calculated (Fig.~\ref{fig:angle}).
Here, the bond distances of Ge--O were defined by the first minimum in the Ge--O partial pair distribution functions.
The O--Ge--O angle distributions are indicative of the Ge-centered structural motifs, while the Ge--O--Ge angle
is indicative of the connection between those structural motifs. Much emphasis has been put on obtaining the Ge--O--Ge angles 
\cite{hussin1999, neuefeind1996} because 
it is a vital parameter in the continuous random network model \cite{zachariasen1932}, which is a widely accepted 
starting point for glass network theory.

Under ambient pressure, the O--Ge--O bond angle distribution has a peak at around \SI{109}{\degree},
which corresponds to the ideal internal angle in the tetrahedron;
this indicates the existence of a dominant tetrahedral structural motif in the amorphous material similar to SiO$_2$ \cite{hussin1999,pandey2015,micoulaut2006}.
Little change is seen in the position of this peak up to \SI{4}{\giga\pascal}, meaning that amorphous GeO$_2$ retains
most of its tetrahedrally-dominated structure within this pressure range.
There is also a shoulder at \SI{120}{\degree} under ambient condition, which would be a signature of the existence of trigonal bipyramids if there are GeO$_5$ units.
However, analysis of the local coordination, as discussed above, revealed that there are no GeO$_5$ units at ambient pressure (Fig.~\ref{fig:coord}). Therefore this shoulder originates from distorted GeO$_4$ units.

Turning our attention to the inter-tetrahedral Ge--O--Ge bond angle, we find that the distribution exhibits a broad peak centered at around \SI{125}{\degree}, which is about \SI{10}{\degree} lower than that reported by several works in the literature \cite{salmon2007, hussin1999, neuefeind1996, wezka2012}. The source of this discrepancy will be discussed later. It is also noted that the peak position is much lower than in amorphous SiO$_2$ whose Si--O--Si peak position is reported to be in the range of $~\SI{140}{\degree}$--$\SI{150}{\degree}$ \cite{hussin1999,pandey2015,micoulaut2006}. This peak shifts to a slightly lower angle at \SI{4}{\giga\pascal}, confirming that compaction proceeds via reorganization of corner-sharing tetrahedral GeO$_4$ units \cite{wezka2012}.


With further pressure application in the range of \SI{5.9}{\giga\pascal} $< P <$ \SI{8.0}{\giga\pascal}, the peak positions shift to lower angles and new peaks around \SI{100}{\degree} and \SI{90}{\degree} emerge for Ge--O--Ge and O--Ge--O bond angle distributions, respectively.
This indicates that compaction proceeds via corner-sharing tetrahedral units converting to edge-sharing polyhedral units.
When $P > $ \SI{11.5}{\giga\pascal}, the lower side peaks show remarkable increase; clear peaks appear in the O--Ge--O angle distribution around \SI{90}{\degree} and \SI{170}{\degree} corresponding to the formation of octahedral GeO$_6$ units.
In the case of amorphous SiO$_2$, very little change in the structure has been detected in this pressure range \cite{micoulaut2006,onodera2020}.


These results are generally in good agreement with the literature, although quantitative comparison is difficult due
to the noisiness in the bond angle distribution (as an example, peaks and shoulders from MD simulation using DIPPIM \cite{wezka2012} are compared to the present RMC results in Fig.~\ref{fig:angle}). However, the discrepancy in the Ge--O--Ge angle at ambient pressure is rather clear;
the present RMC model shows a Ge--O--Ge (inter-tetrahedral) bond angle centered around \SI{125}{\degree}, which is 
about \SI{10}{\degree} lower than the mean bond
angle reported in experimental works \cite{salmon2007, hussin1999, neuefeind1996, wezka2012}.
Those works inferred inter-tetrahedral bond angles from simple trigonometric arguments using peak-positions of Ge--O and Ge--Ge pair distribution functions.
On the other hand, we used the minimum after the first peak of the Ge--O partial pair distribution function as the cutoff and extracted all Ge--O--Ge within that range from the RMC structure model. Then, the angles of the extracted Ge--O--Ge structures were plotted as distributions.
It should be noted that the experimental radial distributions are also reproduced by the present RMC model, which leads to the same bond angle by using the same simple trigonometric argument.
The RMC results are also \SI{10}{\degree} lower than DIPPIM MD results as well as a report that combined classical MD using an SiO$_2$ potential with first-principles relaxation \cite{giacomazzi2005}; we tentatively suggest that this is because the MD potentials were constructed to reproduce the bond angles reported in experimental works.
To confirm the present Ge--O--Ge angle distribution (RMC analysis often result in broader peaks compared to FPMD \cite{akola2013}), we constructed an amorphous GeO$_2$ model from a melt-quench FPMD simulation and obtained a Ge--O--Ge bond angle peak at the same position as the present  RMC result (chained line in Fig.~\ref{fig:angle}).
It should also be noted that a previous FPMD work by a different group also reported a peak around \SI{125}{\degree}.

\subsection{Cavity Analysis}
\begin{figure}[ht]
    \centering
    \includegraphics[width=6.28cm]{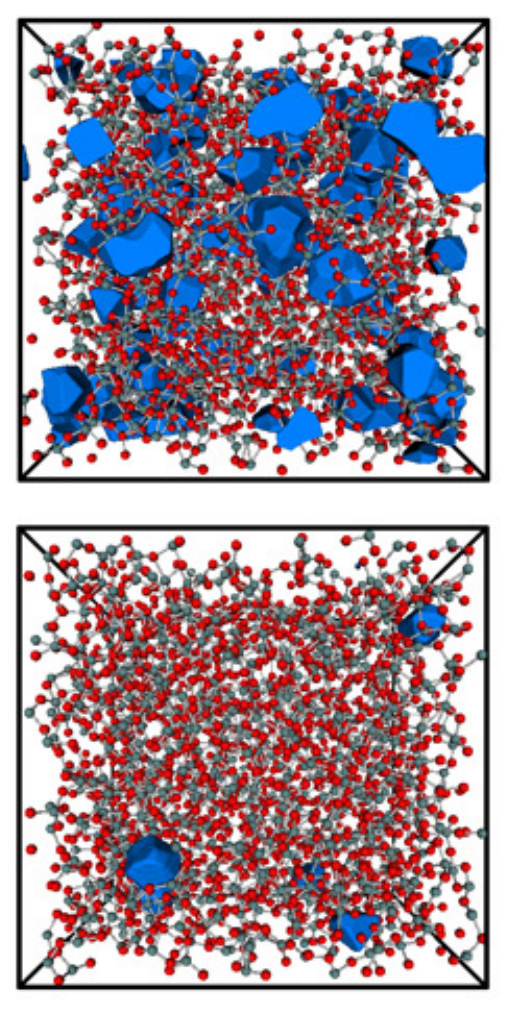}
    \caption{\label{fig:cavity}(Color online) Cavity analysis of the three-dimensional structure obtained by the RMC method.   Cavities comprised \SI{5.7}{\percent} of the total volume under ambient conditions (upper) and \SI{0.1}{\percent} at \SI{4}{\giga\pascal} (bottom).}
\end{figure}
The coordination and bond angle analyses presented above reflect the local structure only up to third nearest neighbor.
To complement them with an indicator of the topological network beyond third nearest neighbor, cavity analysis is useful.
Cavities are known to influence many materials properties, and its importance has been pointed out 
particularly for determining SET/RESET behavior in phase change materials \cite{akola2012a}.
Fig. \ref{fig:cavity} shows cavity volumes calculated and visualized using pyMolDyn program \cite{heimbach2017} with a cutoff radius of \SI{2.8}{\angstrom}; 
these cavities account for \SI{5.7}{\percent} of the total volume under ambient conditions, while it accounts for only \SI{0.1}{\percent} of the volume at \SI{4}{\giga\pascal} $(\rho/\rho_0=1.15)$. 
This clarifies how compaction proceeds at lower pressures without clear changes in the tetrahedral network structure. The rearrangement of the tetrahedral network, which was indicated by the decrease of the Ge--O--Ge bond angle (Fig. \ref{fig:angle}), leads to collapsing of cavity volumes. These behaviors can also be correlated with the changes in partial structure factors (Fig. \ref{fig:p_sq}). 
That is, the disappearance of the FSDP in the O--O partial structure factor is seen at the same time as the annihilation of cavities at \SI{4}{\giga\pascal}.
Once the cavities are mostly annihilated, the network structure cannot withstand further compaction without changes in the local structural motifs.
Kono \textit{et al.} observed pressure-induced behavior of cavities in silica glass by combining molecular dynamics and RMC analysis \cite{kono2022}. 
Their results are also similar to ours, but the magnitude of changes in the partial structure factors are much smaller due to differences in their pressure sensitivities.
It can also be suggested that these cavities may be the source of the ``reversibility window'' observed in silica glass \cite{trachenko2003}, as almost no changes in the bond topology
are seen below \SI{4}{\giga\pascal} and rebirth of these cavities can be expected upon pressure release.

\section{C\lowercase{onclusion}}
In this paper, RMC analysis of the structure of amorphous GeO$_2$ has been performed based on structure factors from NDIS experiments under pressure. The aim was to provide an experiment-based analysis of the local and intermediate-range  structure to complement the literature where some discrepancy exists among classical/{\it ab inito} MD studies and analyses based on experimental data. 
The present RMC results are based solely on the experimental structure factors and no assumptions enter the analysis except for minimal coordination constraints noted above. 
The agreement of RMC results with several MD studies as well as our own FPMD melt-quench simulation thus allows us to confirm the following overall picture of the compaction mechanisms in amorphous GeO$_2$:
\begin{itemize}
    \item The amorphous structure under ambient pressure is comprised predominantly of
    tetrahedral units, and those units are retained up to $\sim$\SI{4}{\giga\pascal}.
    \item The compaction below $\sim$\SI{4}{\giga\pascal} proceeds through collapsing of cavity volumes through
        buckling of the tetrahedral network structure.
    \item Further compaction at above $\sim$\SI{4}{\giga\pascal} after cavities are annihilated require changes in the
    structural motifs: the tetrahedral units are gradually converted to square pyramidal GeO$_5$ units at 
    $\rho/\rho_0 > 1.15$, which are, in turn, converted even more gradually to octahedral GeO$_6$ units at 
    $\rho/\rho_0 > 1.38$.
\end{itemize}
This work demonstrates the necessity of structure modeling in three-dimensional real space. In this regard, the RMC method combined with NDIS has turned out to be an extremely effective approach.
\begin{acknowledgments}
We gratefully thank Philip Salmon for providing his neutron diffraction data.
This research was financially supported by the Japan Society for Promotion of Science (JSPS) KAKENHI Grant Number 16H03903 and 20H02430. Kenta Matsutani and Shusuke Kasamatsu are supported by Japan Science and Technology Agency (JST) FOREST Program (Grant Number JPMJFR2037, Japan).
\end{acknowledgments}


%
%

%


\bibliographystyle{jpsj}
\bibliography{ref}

\begin{thebibliography}{10}

\bibitem{lee2008}
S.~K. Lee, J.-F. Lin, Y.~Q. Cai, N.~Hiraoka, P.~J. Eng, T.~Okuchi, H.-k. Mao, Y.~Meng, M.~Y. Hu, P.~Chow, J.~Shu, B.~Li, H.~Fukui, B.~H. Lee, H.~N. Kim, and C.-S. Yoo: PNAS {\bfseries 105} (2008) 7925.

\bibitem{petitgirard2019}
S.~Petitgirard, C.~Sahle, C.~Weis, K.~Gilmore, G.~Spiekermann, J.~Tse, M.~Wilke, C.~Cavallari, V.~Cerantola, and C.~Sternemann: Geochem. Persp. Let.  (2019) 32.

\bibitem{wilding2006}
M.~C. Wilding, M.~Wilson, and P.~F. McMillan: Chemical Society Reviews {\bfseries 35} (2006) 964.

\bibitem{onodera2020}
Y.~Onodera, S.~Kohara, P.~S. Salmon, A.~Hirata, N.~Nishiyama, S.~Kitani, A.~Zeidler, M.~Shiga, A.~Masuno, H.~Inoue, S.~Tahara, A.~Polidori, H.~E. Fischer, T.~Mori, S.~Kojima, H.~Kawaji, A.~I. Kolesnikov, M.~B. Stone, M.~G. Tucker, M.~T. McDonnell, A.~C. Hannon, Y.~Hiraoka, I.~Obayashi, T.~Nakamura, J.~Akola, Y.~Fujii, K.~Ohara, T.~Taniguchi, and O.~Sakata: NPG Asia Materials {\bfseries 12} (2020) 1.
\newblock Number: 1 Publisher: Nature Publishing Group.

\bibitem{micoulaut_structure_2006}
M.~Micoulaut, L.~Cormier, and G.~S. Henderson: Journal of Physics: Condensed Matter {\bfseries 18} (2006) R753.

\bibitem{itie1989}
J.~P. Itie, A.~Polian, G.~Calas, J.~Petiau, A.~Fontaine, and H.~Tolentino: Phys. Rev. Lett. {\bfseries 63} (1989) 398.

\bibitem{lelong2012}
G.~Lelong, L.~Cormier, G.~Ferlat, V.~Giordano, G.~S. Henderson, A.~Shukla, and G.~Calas: Phys. Rev. B {\bfseries 85} (2012) 134202.

\bibitem{wezka2012}
K.~Wezka, P.~S. Salmon, A.~Zeidler, D.~A.~J. Whittaker, J.~W.~E. Drewitt, S.~Klotz, H.~E. Fischer, and D.~Marrocchelli: J. Phys.: Condens. Matter {\bfseries 24} (2012) 502101.

\bibitem{shanavas2006}
K.~V. Shanavas, N.~Garg, and S.~M. Sharma: Phys. Rev. B {\bfseries 73} (2006) 094120.

\bibitem{li2009}
T.~Li, S.~Huang, and J.~Zhu: Chemical Physics Letters {\bfseries 471} (2009) 253.

\bibitem{zhu2009}
X.~F. Zhu and L.~F. Chen: Physica B: Condensed Matter {\bfseries 404} (2009) 4178.

\bibitem{brazhkin2011}
V.~V. Brazhkin, A.~G. Lyapin, and K.~Trachenko: Phys. Rev. B {\bfseries 83} (2011) 132103.

\bibitem{vaccari2009}
M.~Vaccari, G.~Aquilanti, S.~Pascarelli, and O.~Mathon: J. Phys.: Condens. Matter {\bfseries 21} (2009) 145403.

\bibitem{marrocchelli2009}
D.~Marrocchelli, M.~Salanne, P.~A. Madden, C.~Simon, and P.~Turq: Mol. Phys. {\bfseries 107} (2009) 443.

\bibitem{marrocchelli2010}
D.~Marrocchelli, M.~Salanne, and P.~A. Madden: J. Phys.: Condens. Matter {\bfseries 22} (2010) 152102.

\bibitem{salmon2007}
P.~S. Salmon, A.~C. Barnes, R.~A. Martin, and G.~J. Cuello: J. Phys.: Condens. Matter {\bfseries 19} (2007) 415110.

\bibitem{evrard2005}
G.~Evrard and L.~Pusztai: J. Phys.: Condens. Matter {\bfseries 17} (2005) S1.

\bibitem{leadbetter1972}
A.~J. Leadbetter and A.~C. Wright: Journal of Non-Crystalline Solids {\bfseries 7} (1972) 37.

\bibitem{salmon2012}
P.~S. Salmon, J.~W.~E. Drewitt, D.~A.~J. Whittaker, A.~Zeidler, K.~Wezka, C.~L. Bull, M.~G. Tucker, M.~C. Wilding, M.~Guthrie, and D.~Marrocchelli: J. Phys.: Condens. Matter {\bfseries 24} (2012) 415102.

\bibitem{heimbach2017}
I.~Heimbach, F.~Rhiem, F.~Beule, D.~Knodt, J.~Heinen, and R.~O. Jones: J. Comput. Chem. {\bfseries 38} (2017) 389.

\bibitem{perdew1996}
J.~P. Perdew, K.~Burke, and M.~Ernzerhof: Phys. Rev. Lett. {\bfseries 77} (1996) 3865.

\bibitem{kresse1996}
G.~Kresse and J.~Furthm{\"u}ller: Phys. Rev. B {\bfseries 54} (1996) 11169.

\bibitem{guthrie2004}
M.~Guthrie, C.~A. Tulk, C.~J. Benmore, J.~Xu, J.~L. Yarger, D.~D. Klug, J.~S. Tse, H.-k. Mao, and R.~J. Hemley: Phys. Rev. Lett. {\bfseries 93} (2004) 115502.

\bibitem{sanloup2013}
C.~Sanloup, J.~W.~E. Drewitt, Z.~Kon{\^o}pkov{\'a}, P.~{Dalladay-Simpson}, D.~M. Morton, N.~Rai, W.~{van Westrenen}, and W.~Morgenroth: Nature {\bfseries 503} (2013) 104.

\bibitem{cabaret2007}
D.~Cabaret, F.~Mauri, and G.~S. Henderson: Phys. Rev. B {\bfseries 75} (2007) 184205.

\bibitem{hussin1999}
R.~Hussin, R.~Dupree, and D.~Holland: Journal of Non-Crystalline Solids {\bfseries 246} (1999) 159.

\bibitem{neuefeind1996}
J.~Neuefeind and K.-D. Liss: Berichte Bunsenges. F\"ur Phys. Chem. {\bfseries 100} (1996) 1341.

\bibitem{zachariasen1932}
W.~H. Zachariasen: J. Am. Chem. Soc. {\bfseries 54} (1932) 3841.

\bibitem{pandey2015}
A.~Pandey, P.~Biswas, and D.~A. Drabold: Phys. Rev. B {\bfseries 92} (2015) 155205.

\bibitem{micoulaut2006}
M.~Micoulaut, Y.~Guissani, and B.~Guillot: Phys. Rev. E {\bfseries 73} (2006) 031504.

\bibitem{giacomazzi2005}
L.~Giacomazzi, P.~Umari, and A.~Pasquarello: Phys. Rev. Lett. {\bfseries 95} (2005) 075505.

\bibitem{akola2013}
J.~Akola, S.~Kohara, K.~Ohara, A.~Fujiwara, Y.~Watanabe, A.~Masuno, T.~Usuki, T.~Kubo, A.~Nakahira, K.~Nitta, T.~Uruga, J.~K.~R. Weber, and C.~J. Benmore: PNAS {\bfseries 110} (2013) 10129.

\bibitem{akola2012a}
J.~Akola and R.~O. Jones: Phys. Status Solidi B {\bfseries 249} (2012) 1851.

\bibitem{kono2022}
Y.~Kono, K.~Ohara, N.~M. Kondo, H.~Yamada, S.~Hiroi, F.~Noritake, K.~Nitta, O.~Sekizawa, Y.~Higo, Y.~Tange, H.~Yumoto, T.~Koyama, H.~Yamazaki, Y.~Senba, H.~Ohashi, S.~Goto, I.~Inoue, Y.~Hayashi, K.~Tamasaku, T.~Osaka, J.~Yamada, and M.~Yabashi: Nat Commun {\bfseries 13} (2022) 2292.

\bibitem{trachenko2003}
K.~Trachenko and M.~T. Dove: Phys. Rev. B {\bfseries 67} (2003) 212203.

\end{thebibliography}
\end{document}